\documentclass[aps,prb,twocolumn,showpacs,superscriptaddress,preprintnumbers,amsmath,amssymb]{revtex4-1}

\usepackage{graphicx}
\usepackage{hyperref}
\usepackage{epsfig}
\usepackage{dcolumn}
\usepackage{bm}
\usepackage{longtable}
\usepackage{color}
\pdfoutput=1

\begin{document}


\title{Small and nearly isotropic hole-like Fermi surfaces in LiFeAs detected through de Haas van Alphen-effect}



\author{B.\ Zeng}
\affiliation{National High Magnetic Field Laboratory, Florida
State University, Tallahassee-FL 32310, USA}
\author{D.\ Watanabe}
\affiliation{Department of Physics, Kyoto University, Kyoto 606-8502, Japan}
\affiliation{National High Magnetic Field Laboratory, Florida
State University, Tallahassee-FL 32310, USA}
\author{Q.\ R.\ Zhang}
\affiliation{National High Magnetic Field Laboratory, Florida
State University, Tallahassee-FL 32310, USA}
\author{G.\ Li}
\affiliation{National High Magnetic Field Laboratory, Florida
State University, Tallahassee-FL 32310, USA}
\author{T. Besara}
\affiliation{National High Magnetic Field Laboratory, Florida
State University, Tallahassee-FL 32310, USA}
\author{T. Siegrist}
\affiliation{National High Magnetic Field Laboratory, Florida
State University, Tallahassee-FL 32310, USA}
\affiliation{Department of Chemical and Biomedical Engineering, Florida State University, Tallahassee, Florida 32310, USA.}
\author{L. Y. Xing}
\affiliation{Institute of Physics, Chinese Academy of Sciences, Beijing 100190, China }
\author{X. C. Wang}
\affiliation{Institute of Physics, Chinese Academy of Sciences, Beijing 100190, China }
\author{C. Q. Jin}
\affiliation{Institute of Physics, Chinese Academy of Sciences, Beijing 100190, China }
\author{P. Goswami}
\affiliation{National High Magnetic Field Laboratory, Florida
State University, Tallahassee-FL 32310, USA}
\author{M. D. Johannes}
\affiliation{Center for Computational Materials Science, Naval Research Laboratory, Washington, DC 20375, USA}
\author{L.\ Balicas} \email{balicas@magnet.fsu.edu}
\affiliation{National High Magnetic Field Laboratory, Florida
State University, Tallahassee-FL 32310, USA}


\date{\today}

\begin{abstract} LiFeAs is unique among the arsenic based Fe-pnictide superconductors
because it is the only nearly stoichiometric compound which does \emph{not} exhibit magnetic
order. This is at odds with electronic structure calculations which find a very stable
magnetic state and predict cylindrical hole- and electron-like Fermi surface sheets whose
geometry suggests spin fluctuations and a possible instability toward long-range ordering at
the nesting vector\cite{singh, mazin}. In fact, a complex magnetic phase-diagram is indeed
observed in the isostructural NaFeAs compound \cite{parker}. Previous angle resolved
photoemission (ARPES) experiments \cite{borisenko} revealed the
existence of both hole and electron-like surfaces, but with rather
distinct cross-sectional areas and an absence of the nesting that is
thought to underpin both magnetic order and superconductivity in the pnictide family of superconductors.
These ARPES observations were challenged by subsequent de Haas van Alphen (dHvA) measurements which detected a few, electron like
Fermi surface sheets in rough agreement with the original band calculations \cite{putzke}.
Here, we show a detailed dHvA study unveiling additional, small and nearly isotropic Fermi
surface sheets in LiFeAs single crystals, which ought to correspond to hole-like orbits, as
previously observed by ARPES \cite{borisenko}. Therefore, our results conciliate the apparent discrepancy between ARPES \cite{borisenko} and the previous dHvA results \cite{putzke}.
The small size of these Fermi surface pockets suggests a prominent role for
the electronic correlations in LiFeAs \cite{haule1, yin, ferber, haule}. The absence of gap nodes, in combination with
the coexistence of quasi-two-dimensional and three-dimensional Fermi surfaces, favor
a $s$-wave pairing symmetry for LiFeAs. But similar electron-like Fermi surfaces combined with very different hole
pockets between LiFeAs and LiFeP, suggest that the nodes in the gap function of
LiFeP \cite{hashimoto} might be located on the hole-pockets. This would be difficult to conciliate with the current understanding of the $s\pm$-scenario \cite{mazin}.
\end{abstract}


\maketitle

\section{Introduction}
The uniqueness of LiFeAs among the Fe pnictides, stems from the simple
fact that it is a nearly stoichiometric superconductor: previous studies
indicate a molar ratio for Li/Fe/As of 0.99:1.00:1.00, through
inductively coupled plasma mass spectrometry \cite{chu, morozov, wang}.
In contrast to most Fe-based superconductors, in LiFeAs very small
deviations in stoichiometry such as excess Fe, suppresses
superconductivity very quickly: LiFe$_{1.01}$As would no longer be
superconducting according to the detailed structural analysis of Refs.
\onlinecite{pitcher, wright}.

Our band structure calculations, like earlier ones\cite{singh, mazin}, indicate the existence of three
hole-like Fermi surface sheets in LiFeAs, i.e. a small elongated ellipsoid and two corrugated cylinders
around the $\Gamma$-point of the Brillouin-zone, as well as two electron-like corrugated cylinders
around its $M$-point, see Fig. 1. As with the parent compounds of other Fe-based superconductors, the
calculated magnetic ground state is energetically favored over a non-magnetic one, and the calculated
cross-sectional areas of the large hole- and electron-like cylinders are very similar.  In combination,
these features point to a magnetic instability that must be suppressed (in general by doping or
pressure) before the superconducting state can emerge. Instead, LiFeAs has a superconducting ground
state with $T_c$ $\sim 18$ K, a situation comparable to LiFeP which has a remarkably similar Fermiology
and $T_c = 5$ K.

\begin{figure}[htbp]
\begin{center}
\includegraphics[width = 8.6 cm]{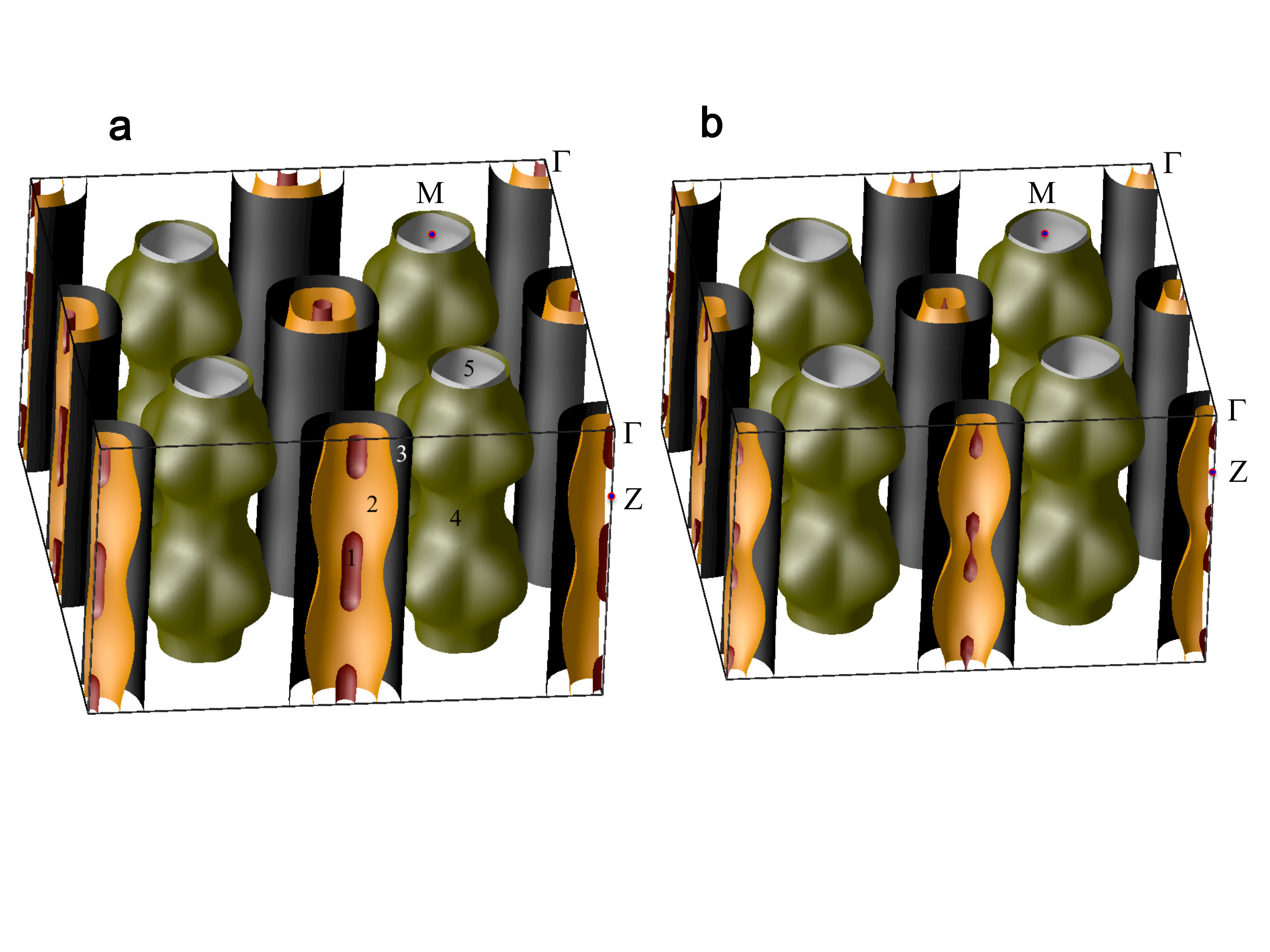}
\caption{(Color online) \textbf{a} Fermi surface of LiFeAs according Density Functional Theory calculations. \textbf{b} Same as in \textbf{a} after a shift of the Fermi level. Notice how the
the hole-like Fermi surface sheets around the $\Gamma$ point shrink in size when one increases the electron count.}
\end{center}
\end{figure}
Furthermore, although a number of experiments indicate a fully gapped as well as multi-gapped superconducting state for LiFeAs \cite{wei, stockert, kim, tanatar, hashimoto, umezawa}, penetration depth measurements
unveil nodal lines for the gap function of LiFeP \cite{hashimoto}. Earlier angle resolved photoemission experiments (ARPES) \cite{borisenko}, revealed the existence of two-hole like cylinders at the $\Gamma$ point,
with quite distinct cross-sectional areas with respect to those of the electron sheets at the $M$-point. This weak ``nesting" between hole- and electron-like Fermi surface sheets would explain the absence of antiferromagnetism,
but instead generate antiferromagnetic fluctuations, which is a possible ingredient for the superconducting pairing mechanism. Subsequent ARPES studies \cite{umezawa,haule} confirmed these observations, and unveiled
fully gapped superconducting gaps in all four FS sheets, with some of the gaps displaying a strong anisotropy in $k$-space \cite{umezawa}.
\begin{figure*}[htbp]
\begin{center}
\includegraphics[width = 14 cm]{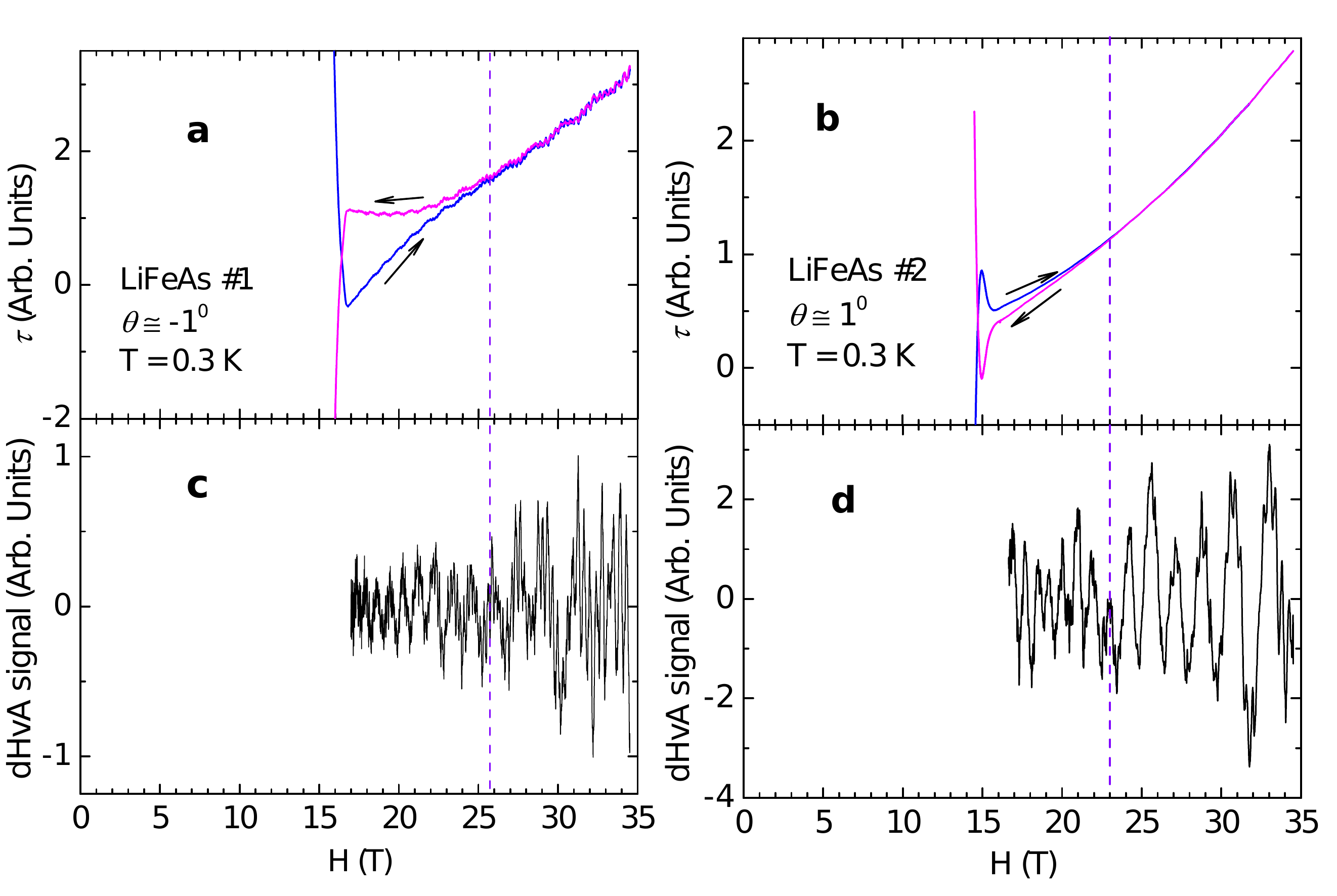}
\caption{(Color online) \textbf{a} Magnetic torque $\tau$ as a function of the magnetic field $H$ for a LiFeAs single-crystal (crystal $\sharp 1$) at a temperature $T = 0.3$ K and an angle $\theta \simeq 1^{\circ}$
between $H$ and the inter-planar \emph{c}-axis. Blue and magenta lines indicate increasing and decreasing field sweeps.
Notice how the large (and incomplete) hysteresis loop is followed at higher fields by a smaller one as previously seen in Ref. \onlinecite{li}.
\textbf{b} Same as in \textbf{a} but for a second LiFeAs single-crystal (crystal $\sharp 2$). \textbf{c} The oscillatory component,
or the de Haas van Alphen-effect, superimposed onto the $\tau (H)$ trace shown in \textbf{a} after the subtraction of a polynomial-fit, and as a function of $H$.
Vertical purple line indicates the irreversibility field. Hence, the quantum oscillatory phenomena is observed deep inside the irreversible superconducting region.
\textbf{d} Same as in \textbf{c} but for the torque data in \textbf{b}. Notice how the irreversibility field is higher for sample $\sharp 1$ when compared to $\sharp 2$, therefore indicating a higher sample quality.}
\end{center}
\end{figure*}

Therefore, it came as a surprise that a subsequent investigation on de Haas van Alphen-effect, on
both LiFeP and LiFeAs, found a general good agreement with Density Functional Theory (DFT)
calculations \cite{putzke}. Although, in LiFeAs only three of the possible ten orbits were
observed, and attributed to orbits on the electron-like sheets. To address this apparent
discrepancy concerning the size of the hole-like Fermi surfaces, a couple of reports \cite{ferber,
haule} performed a Dynamical Mean Field Theory (DMFT) study, to introduce the effect of electronic
correlations on the band structure and on the concomitant Fermi surfaces resulting from the DFT
calculations. Several studies \cite{haule1, yin, ferber, haule}, indicate that the correlations are mainly controlled
by the value of the Hund$^{\prime}$s rule coupling $J$.  Hund$^{\prime}$s coupling was found to shrink the middle
hole-pocket  which has $t_{2g}$, $d_{xz}$ and $d_{yz}$ orbital character, \emph{leaving the electron
FS sheets intact} \cite{ferber}. The fact that correlations tend to weaken, or even suppress, the nesting between
FSs in LiFeAs, is confirmed by the DFT-DMFT study of Ref. \onlinecite{haule}. It finds that
correlations affect the hybridization magnitude between the orbitals, resulting in a net transfer
of charges from the $d_{xy}$ to the $d_{xz}/d_{yz}$ orbitals. This results in a downshift of the
$d_{xz}/d_{yz}$-bands and in an upshift of $d_{xy}$-related bands with a concomitant decrease
(increase) in the size of the hole FSs having $d_{xz}/d_{yz}$ ($d_{xy}$) -character. Similarly,
Ref. \onlinecite{haule} finds practically no effect of the correlations on the geometry of the
electron-like FSs.

Here, we unveil a study on the de Haas van Alphen-effect (dHvA) on
LiFeAs single crystals revealing a series of small extremal Fermi
surface cross-sectional areas, in addition to the previously observed
electron-like ones \cite{putzke}. These orbits can only be attributed to
the two inner hole-like surfaces, not previously seen by dHvA. We find
evidence for an orbit whose area matches the DFT prediction for the
innermost hole-pocket. However, and surprisingly, it is observed to be
far less anisotropic than the DFT prediction, seen in Fig. 1. We also
detect another set of smaller frequencies, i.e. 330 and 460 T, which can
only correspond to the predicted, middle hole-like Fermi surface around
the $\Gamma$-point. An estimate of the cross-sectional area for this FS
based on previous ARPES measurements \cite{borisenko}, yields a value of
$3.215 \times 10^{-3}$ \AA$^{-2}$ which is nearly perfectly matched by
the Onsager cross-sectional area $(A = 2 \pi e F/ \hbar )$ related to
the frequency $F= 330$ T, or $3.15 \times 10^{-3}$ \AA$^{-2}$.
Therefore, and in agreement with the ARPES results, its size is
considerably smaller than the DFT prediction implying that the
electronic structure of LiFeAs is severely affected by electronic
correlations (following the above DMFT arguments). Even more remarkable
is the fact that these orbits are also nearly three-dimensional. In
general, the DFT approximation to the exchange correlation functional
predicts metallic bandwidths that are larger than experiment
\cite{paier}, yielding greater isotropy in Fermi surfaces. It is
therefore intriguing that our experimental measurements show both a strong
renormalization of the calculated electron masses (bandwidths), and
simultaneously an increase in isotropy. We conclude that our results suggest that the
nodes in the gap function of LiFeP \cite{hashimoto} might be located on
the hole-pockets, given that the electron-like FSs of LiFeAs and LiFeP
are very similar in size and geometry, but the size of the hole-pockets
in LiFeAs are considerably smaller. Coupled to nearly isotropic Fermi
surface sheets, this would also suggest a pairing symmetry
possibly distinct from the originally proposed $s^{\pm}$ scenario for
Fe pnictide superconductors \cite{mazin}.

\section{Experimental Results}

High-quality single crystals of LiFeAs have been grown by a self-flux technique \cite{xcwang1, xcwang2}.
The precursor of Li$_3$As was synthesized from Li pieces and As chips that were sealed in a Nb tube under Ar atmosphere and then treated at 650 $^{\circ}$C for 15 h in a sealed quartz tube.
The Li$_3$As, Fe, and As powders were mixed using the following ratio: Li:Fe:As = 1:0.3:1. The powder mixture was then pressed into a pellet in an alumina oxide tube.
The sealed quartz tube was heated at 800 $^{\circ}$C for 10 h before heating up to 1100 $^{\circ}$C at which it was held for another 10 h.
Finally, it was cooled down to 800 $^{\circ}$C at a rate of 2 $^{\circ}$C per hour. Crystals with a size up to 4 mm $\times$ 3 mm $\times$ 0.5 mm were obtained.
The whole preparation work was performed within a glove box (or in high purity Ar gas). Torque magnetometry was measured by using the capacitive method with a 0.0762 mm thick CuBe lever.
The angle between the sample and the external field was measured with a Hall probe.

Figure 2 shows the magnetic torque $\tau$ as a function of the magnetic-field (as well as the oscillatory component superimposed onto it, or the dHvA-effect) for two LiFeAs single-crystals named here thereafter as crystals $\sharp$ 1 and $\sharp$ 2, respectively at $T = 0.5$ K and $T = 0.3$ K and for fields nearly aligned along the \emph{c}-axis. Blue line represents increasing field scan, while the magenta line represents the decreasing field one. The large hysteresis corresponds to the superconducting signal, which is dominated by vortex pinning, and is several orders of magnitude larger ($> 10^3$) than the signal in provenance of the paramagnetic metallic state. Notice how the original (incomplete) hysteresis loop is followed by a much smaller one, where the lower branch becomes the upper one, and the upper one becomes the lower branch. Such a behavior, suggesting a crossover from a net diamagnetic response due to vortex pinning, towards an enhanced paramagnetic-like and hysteretic response (\emph{within the superconducting state}), was already reported by us in Ref. \onlinecite{li} and attributed to a possible field-induced chiral superconducting state in LiFeAs. We emphasize that we studied well over a dozen crystals, and as clearly seen here, this anomalous hysteretic response is only observed on those crystals with enough purity to display the de Haas van Alphen effect. The quality of the crystal can be judged, for example, through the value of the irreversibility field $H_{\text{irr}}$ (or the value in field where the hysteresis loop closes), which for crystal $\sharp 1$ is $ \sim 26$ T, i.e. considerably higher than the values of $H_{c2}$ previously reported for LiFeAs at this temperature and for this field orientation \cite{khim, cho, kurita}. Notice also that the dHvA-effect is observed deep inside the irreversible superconducting region, i.e from $H \lesssim 17$ T up to $H_{\text{irr}} \sim 26$ T.
\begin{figure*}[htbp]
\begin{center}
\includegraphics[width = 14 cm]{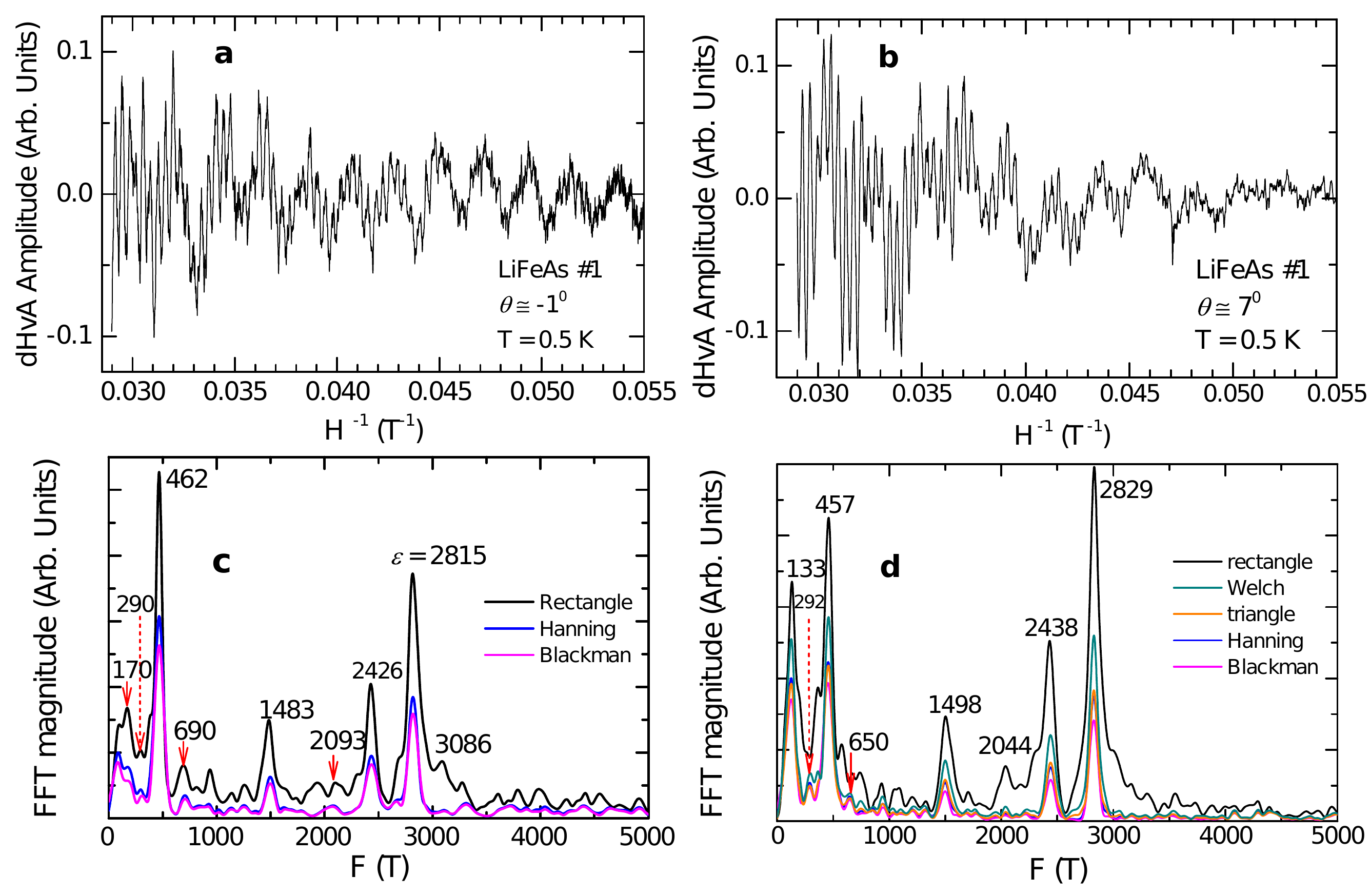}
\caption{(Color online) \textbf{a} Oscillatory component superimposed onto the torque signal obtained at $T = 0.5$ K from crystal $\sharp 1$ as a function of the inverse magnetic field $H^{-1}$ and for an angle $\theta \simeq - 1^{\circ}$ between $H$ and the inter-layer \emph{c}-axis. \textbf{b} Same as in \textbf{a} but for an angle $\theta = 7^{\circ}$. Notice how the position of the main peaks in frequency is independent of the selected spectral window. \textbf{c} Fast Fourier transform of the oscillatory signal shown in \textbf{a} for several spectral windows.  \textbf{d} Same as in \textbf{c} but for $\theta = 7^{\circ}$.}
\end{center}
\end{figure*}
\begin{figure*}[htbp]
\begin{center}
\includegraphics[width = 14 cm]{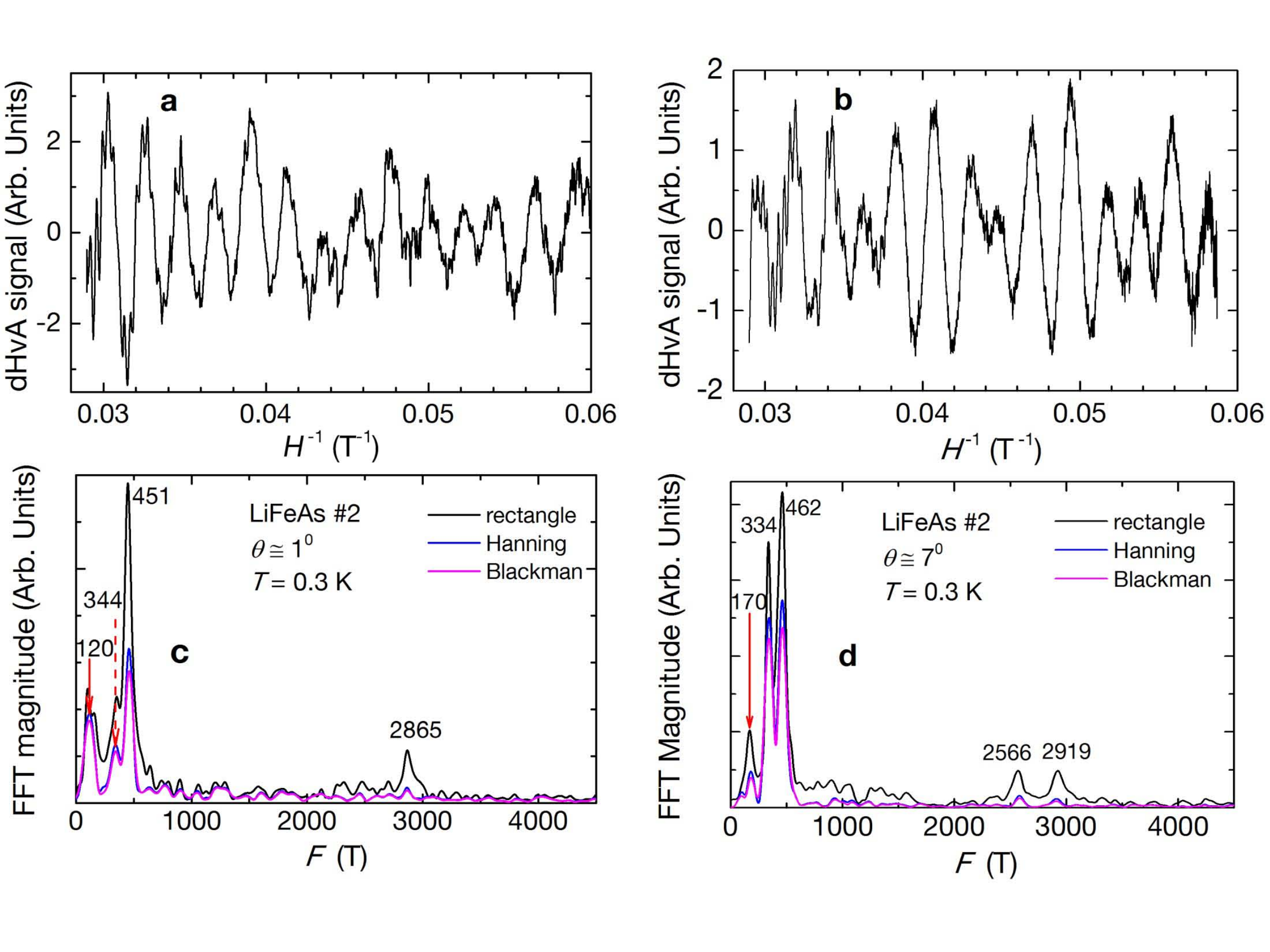}
\caption{Color online) \textbf{a} Oscillatory component superimposed onto the torque signal obtained from crystal $\sharp 2$ at $T = 0.3$ K as a function of the inverse magnetic field $H^{-1}$ and for an angle $\theta \simeq 1^{\circ}$. \textbf{b} Same as in \textbf{a} but for an angle $\theta = 7^{\circ}$. \textbf{c} Fast Fourier transform of the oscillatory signal shown in \textbf{a} for several spectral windows. Same as in \textbf{c} but for $\theta = 7^{\circ}$.}
\end{center}
\end{figure*}

Figure 3 shows the oscillatory signal, or the dHvA-effect, superimposed onto the torque traces acquired at $T = 0.5 $ K for
 LiFeAs single-crystal $\sharp 1$ as a function of the inverse field $H^{-1}$ and for two values of the angle $\theta = -1^{\circ} $ and $ 7^{\circ}$, respectively.
 The dHvA signal is obtained from $\tau(H,T)$ after subtraction of a polynomial background. Fig. 3 also shows the fast Fourier transform (FFT) for each trace, and for several spectral windows.
 Notice how the amplitude of certain peaks observed when using a rectangular spectral window (2093 and 2044 T, or 690 and 650 T in Figs. 3 \textbf{c} and \textbf{d}, respectively) disappear towards
 the level of the background signal when different spectral windows are used. This indicates that they are not intrinsic frequencies but an artifact resulting from spectral leakage.
 Regardless of the chosen spectral window, two sharp peaks are observed at $F_{\delta} \simeq 2400$ T and $F_{\epsilon} \simeq 2800$ T which were previously reported in Ref. \onlinecite{putzke} and
 attributed to orbits on both electron-like Fermi surface sheets: neck and ``belly" cyclotron orbits respectively on the gray and green FS sheets shown in Fig. 1 \textbf{a}.
 We also observed a pronounced, spectral window independent peak around $F \simeq 460$ T which is preceded by a smaller peak at $F \simeq 290$ T.
 These values are relatively close to frequencies of the $\alpha$ orbits predicted by the DFT calculations for the innermost hole-like FS sheet in LiFeP,
 but are clearly at odds with the DFT calculations for LiFeAs. Given their small size, one must assume that these frequencies correspond to orbits on hole FSs.
 We also detect evidence for another pair of smaller frequencies having very close values, respectively $ \sim 130$ and 170 T which coincidentally are very close to the orbits/frequencies
 predicted by DFT for the innermost hole-like FS of LiFeAs, i.e. 121 and 132 T. These values should be taken with a grain of salt, given that the amplitude and even the precise value of such small frequencies,
 can be easily affected by the background subtraction. Finally, we observe another peak at $F = 1500$ T which one could be tempted to attribute to the electron-like $F_{\beta} = 1590$ T orbit reported in Ref. \onlinecite{putzke}.

Given the discrepancy between the DFT calculations and the cross-sectional areas observed by us in
sample $\sharp 1$, it is important to verify the reproducibility of these orbits/frequencies on a
second single-crystal. Therefore, Fig. 4 shows the oscillatory signal superimposed onto the torque
measurements performed on a second crystal (LiFeAs $\sharp 2$) as a function of $H^{-1}$ for $T = 0.3$
K, and for two angles, i.e. $\theta = 1^{\circ}$ and $7^{\circ}$, respectively. The same Fig. 3 also
shows the respective FFTs for several spectral windows, reproducing all the frequencies/orbits (within
a margin of 10 \%) previously observed in crystal $\sharp 1$ except for the frequency $F \simeq 1500$
T. Consequently, one can state with confidence that these small frequencies are intrinsic to LiFeAs,
and given their small size, they ought to correspond to small hole-pockets not seen in the previous
dHvA study \cite{putzke}. As stated earlier, if one just takes the small inner hole-orbit observed by
ARPES, which is depicted in Fig. 1 of Ref. \onlinecite{knolle}, one obtains a rough estimate for its
area of just $\sim 0.12$ \% of the area of the first-Brillouin zone in the $k_x$$k_y$-plane. It is
equivalent to a frequency $F \simeq 338$ T when using the Onsager relation. This frequency is
remarkably close to the position of one of the main peaks seen in Figs. 4 \textbf{c} and \textbf{d} at
$F$ of $\sim 330$ to $\sim 345$ T, respectively. The agreement between ARPES and our dHvA results is
consistent with DFT calculations predicting a neutral surface for LiFeAs after cleaving, and the
absence of a distinct electronic structure at its surface \cite{lankau}. A summary of our dHvA results
as well as a comparison with the DFT calculations is presented in table I. We have chosen not to
include the orbit yielding $ F = 1500$ T observed from crystal $\sharp 1$ because it was not
reproduced by the measurements on crystal $\sharp 2$. Notice the remarkable quasi-particle mass
enhancement for the innermost hole orbits.

\vspace{0.5cm}
\begin{table}
\begin{center}
\begin{tabular}{|ccc|cccc|}
  \hline
  \hline
   & DFT &  &  & Experiment &  &  \\   \hline
  Orbit & $F(T)$ & $m_b$ & Orbit & $F(T)$ & $m^{\star}$ & $\frac{m^{\star}}{m_b}-1$ \\
  $1_a$ & 121 & -0.33 & $\alpha_a$ & 120  & $(1.8 \pm 0.5)$ & 4.45 \\
  $1_b$ & 132 & -0.27 & $\alpha_b$ & 170 & $(1.5 \pm 0.3)$&  3.7 \\
  $2_a$ & 1561 & -2.24 & $\zeta_a$ & 330 & $(4.1 \pm 0.5)$ &  0.83\\
  $2_b$ & 2477 & -1.55 & $\zeta_b$ & 460 & $(0.8 \pm 0.1)$ & -0.48 \\
  $3_a$ & 4597 & -2.18 &  &  &  &  \\
  $4_a$ & 2392 & 1.19 & $\delta$ & 2400 & $(4.4 \pm 0.7)$ & 2.7 \\
  $4_b$ & 6300 & 2.39 &  &  &  &  \\
  $5_a$ & 1590 & 1.58 & $\beta^{\sharp}$ &  &  &  \\
  $5_b$ & 2974 & 0.99 & $\epsilon$ & 2800 & $(4.4 \pm 0.5)$ & 3.44 \\
  \hline
  \hline
\end{tabular}
\caption{Fermi surface orbits, related dHvA frequencies and band masses as predicted by the DFT calculations, compared to the experimentally observed dHvA frequencies, effective masses $m^{\star}$ and effective mass enhancements
$\lambda = \frac{m^{\star}}{m_b}-1$. }
\end{center}
\end{table}

\begin{figure*}[htbp]
\begin{center}
\includegraphics[width = 15 cm]{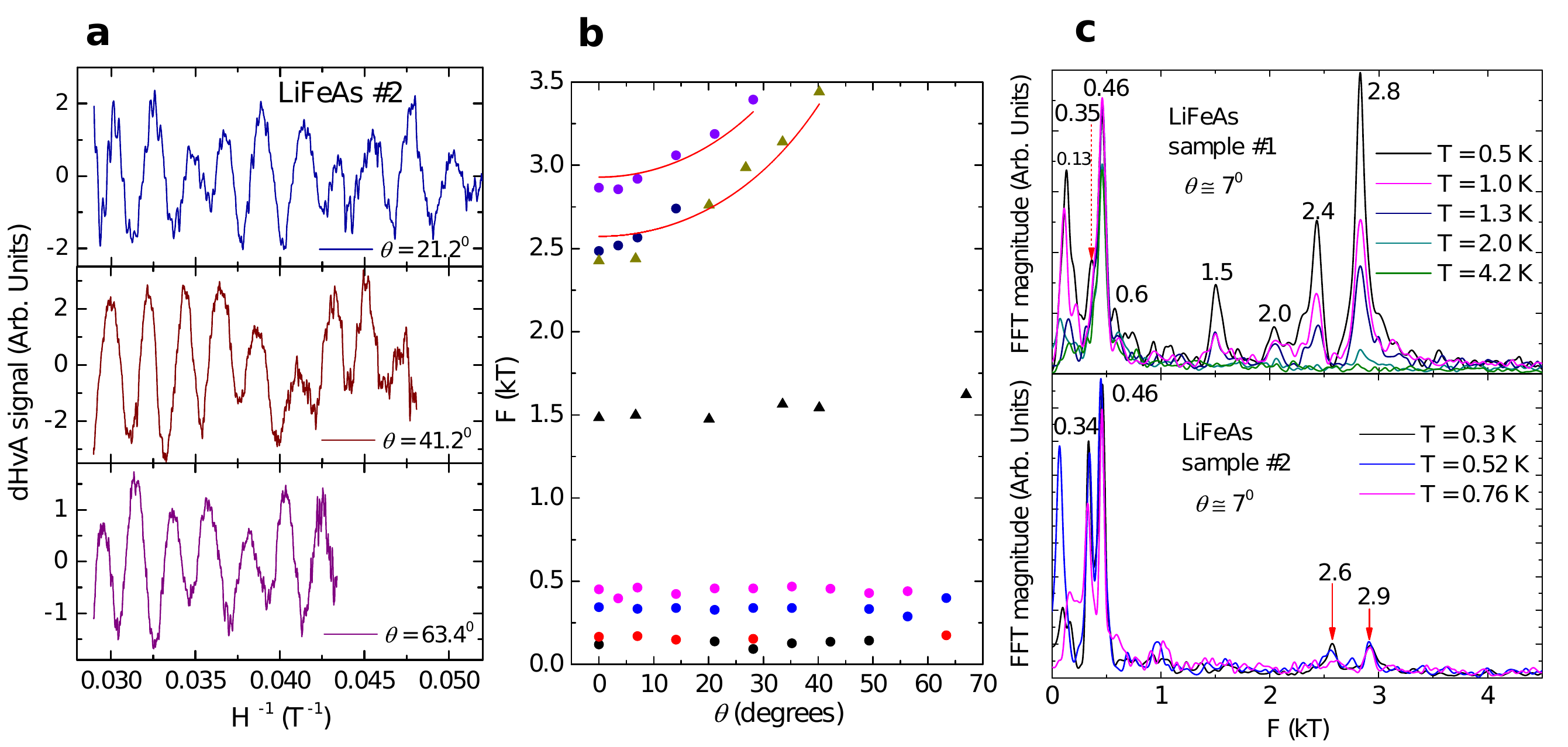}
\caption{\textbf{a} Oscillatory component superimposed onto the torque signal obtained from sample $\sharp 2$, at $T = 0.3$ K, and for several angles $\theta$. \textbf{b} Angular dependence of several of the frequencies observed in the FFT spectra. Red lines are fits to $F = F_0/ \cos (\theta)$, i.e. the angular dependence expected for two dimensional cross-sectional areas. Notice how the smaller frequencies are nearly independent on angle. Therefore, open two-dimensional Fermi surface sheets coexist with closed and nearly isotropic sheets in LiFeAs. Notice that the orbit associated with the frequency $F \simeq 1500$ T observed only in sample $\sharp 1$ (black markers) would also be nearly isotropic. \textbf{c} Magnitude of the fast Fourier transform of the dHvA signal as a function of the frequency $F$, for several temperatures and for samples $\sharp 1$ and $\sharp 2$, respectively. Here, a rectangular spectral window was used.}
\end{center}
\end{figure*}

Figure 5 \textbf{a} shows the oscillatory signal superimposed on the torque response of crystal
$\sharp 2$ at $T = 0.3$ K as a function of $H^{-1}$ and for several angles $\theta$. As seen, at
larger angles the dHvA signal is dominated by the small frequencies. The irreversibility field
increases quickly as the field is oriented towards the \emph{ab}-plane, limiting our ability to access
the metallic state and therefore the dHvA signal for angles beyond $\theta \gtrsim 70^{\circ}$. In
Fig. 5 \textbf{b} we displays the angular dependence of the dHvA frequencies as extracted from both
samples. As expected, the frequencies $F_{\delta} = 2400$ T and $F_{\epsilon} = 2800$ T, attributed to
electron-like Fermi surfaces, follow the $F = F_0/ \cos(\theta)$ dependence (red lines) expected for
two-dimensional orbits. Remarkably, the smaller orbits such as $\alpha_a$ or $\zeta_a$ do not display
much angular dependence at all, implying that these are closed, nearly isotropic orbits. The $\alpha$
orbits were predicted to be three-dimensional but very anisotropic, see Fig. 1 and Refs.
\onlinecite{singh, putzke}. On the contrary, the $\zeta$ orbits were predicted to be two-dimensional.
Perhaps, this discrepancy between experiment and calculations might be addressed by shifting the Fermi
level in the DFT calculations, which as seen in Fig. 1 \textbf{b} would lead to smaller hole-like
Fermi surfaces, and combining it with electronic correlations as in the DMFT approach of Refs.
\onlinecite{ferber, haule}. Finally, Fig. 5 \textbf{c} shows the magnitude of the FFT spectra for both
crystals and for several temperatures at an angle $\theta \simeq 7^{\circ}$. From these traces we
obtained the magnitude of each peak in the FFT spectra as a function of the temperature and extracted
the effective masses (which are shown in Table I) through the Lifshitz-Kosevich formula: $x/\sinh x $
with $x= 14.69 m^{\star}T/H$ and $m^{\star}$ as the effective mass in units of the free electron mass.

\section{Discussion}

LiFeAs was claimed to be placed on the electron overdoped regime
\cite{wang} due to series of observations: i) an extreme sensitivity of
the superconducting state with respect to Li deficiencies which play a
role akin to Zn impurities \cite{yuke} in the overdoped regime of
LaFeAsO$_{1-x}$F$_x$, ii) low-energy, transverse incommensurate spin
excitations by neutron scattering experiments \cite{wang} and iii) a
very asymmetric line in the tunneling spectra \cite{hanke, chi}, similar
to what is seen in electron overdoped NaFe$_{1-x}$Co$_x$As and in sharp
contrast to optimally doped NaFe$_{1-x}$Co$_x$As \cite{xiaodong}. If
this were indeed the case, according to the $s\pm$ scenario
\cite{mazin,chubukov}, one could expect LiFeAs to either display nodes,
or at least a strong angular modulation in the superconducting gap
function on the electron FSs. In effect, for electron overdoped
Ba(Fe$_{1-x}$Co$_x$)$_2$As$_2$ both thermal conductivity \cite{reid} and
heat capacity measurements \cite{jang} indeed reveal evidence for nodes
in its gap function. Nevertheless, all the experimental evidence
collected so far from LiFeAs points towards a fully gapped
superconducting state, albeit with strong gap anisotropies
\cite{hashimoto,umezawa}. Remarkably, evidence for a nodal
superconducting state was instead found in LiFeP which according to the
experimental results shown here, and in agreement with other groups
\cite{borisenko, putzke, hashimoto,umezawa}, has electron-like Fermi
surfaces of nearly the same size as LiFeAs. Furthermore, according to
our work and in agreement with ARPES, the main effect of replacing As
with P is to severely shrink the middle hole FS and to expand the outer
one. If one assumed the same pairing symmetry for both compounds, a
natural assumption given the similar structure and chemical composition,
the nodes in the superconducting gap function would have to be located
on the hole-like Fermi surfaces of LiFeP. It remains to be clarified whether
this fact, coupled with the existence of two sets of nearly isotropic,
three-dimensional hole-like Fermi surfaces in LiFeAs, can be reconciled
with the $s\pm$ pairing symmetry claimed to address nearly all aspects
of the superconducting state in the Fe pnitide/chalcogenide
superconductors \cite{chubukov, yu, chubukov2}. Although, in LiFeAs the absence of nodes and the
co-existence of three-dimensional with two-dimensional Fermi surface sheets, indicates a superconducting pairing
symmetry with a sizeable $s$ component.

\section{Acknowlegements}
We acknowledge useful discussions with A. Chubukov, Z. P. Yin, K. Haule, and P. J. Hirschfeld.
L.~B. is supported by DOE-BES through award DE-SC0002613. T. B. and T. S. are supported by DOE-BES through award DE-SC0008832, and by FSU.
Funding for M.D.J. was provided by the Office of Naval Research (ONR) through the Naval Research Laboratory's Basic Research Program.
The NHMFL is supported by NSF through NSF-DMR-0084173 and the State of Florida.

\end{document}